\documentclass[graybox]{svmult}

\usepackage{mathptmx}       % selects Times Roman as basic font
\usepackage{helvet}         % selects Helvetica as sans-serif font
\usepackage{courier}        % selects Courier as typewriter font
\usepackage{type1cm}        % activate if the above 3 fonts are
\usepackage{makeidx}         % allows index generation
\usepackage{graphicx}        % standard LaTeX graphics tool
\usepackage{multicol}        % used for the two-column index
\usepackage[bottom]{footmisc}% places footnotes at page bottom
\usepackage{amsmath}
\usepackage{multirow}

\makeindex             % used for the subject index
\newcommand{\beq}{\begin{equation}}
\newcommand{\eeq}{\end{equation}}
\newcommand{\beqa}{\begin{eqnarray}}
\newcommand{\eeqa}{\end{eqnarray}}
\newcommand{\ket}[1]{| #1 \rangle}

\newcommand{\mathbb}{\textbf}

\begin{document}

\title*{An introduction to multi-player, multi-choice quantum
games}

% Use \titlerunning{Short Title} for an abbreviated version of
% your contribution title if the original one is too long
\author{Puya Sharif and Hoshang Heydari}
% Use \authorrunning{Short Title} for an abbreviated version of
% your contribution title if the original one is too long
\institute{Puya Sharif \at Department of Physics, Stockholm University, 10691 Stockholm, Sweden. \newline \email{ps@puyasharif.net}
\and Hoshang Heydari \at Department of Physics, Stockholm University, 10691 Stockholm, Sweden. \newline \email{hoshang@fysik.su.se}}
%
% Use the package "url.sty" to avoid
% problems with special characters
% used in your e-mail or web address
%
\maketitle

\abstract*{We give a self contained introduction to a few quantum game protocols, starting with the quantum version of the two-player two-choice game of Prisoners dilemma, followed by a n-player generalization trough the quantum minority games, and finishing with a contribution towards a n-player m-choice generalization with a quantum version of a three-player Kolkata restaurant problem. We have omitted some technical details accompanying these protocols, and instead laid the focus on presenting some general aspects of the field as a whole. This review contains an introduction to the formalism of quantum information theory, as well as to important game theoretical concepts, and is aimed to work as an introduction suiting economists and game theorists with limited knowledge of quantum physics as well as to physicists with limited knowledge of game theory.}

\abstract{We give a self contained introduction to a few quantum game protocols, starting with the quantum version of the two-player two-choice game of Prisoners dilemma, followed by a n-player generalization trough the quantum minority games, and finishing with a contribution towards a n-player m-choice generalization with a quantum version of a three-player Kolkata restaurant problem. We have omitted some technical details accompanying these protocols, and instead laid the focus on presenting some general aspects of the field as a whole. This review contains an introduction to the formalism of quantum information theory, as well as to important game theoretical concepts, and is aimed to work as an introduction suiting economists and game theorists with limited knowledge of quantum physics as well as to physicists with limited knowledge of game theory.}

\section{Introduction}
Quantum game theory is the natural intersection between three fields. Quantum mechanics, information theory and game theory. At the center of this intersection stands one of the most brilliant minds of the 20:th century, John von Neumann. As one of the early pioneers of quantum theory, he made major contributions to the mathematical foundation of the field, many of them later becoming core concepts in the merger between quantum theory and information theory, giving birth to quantum computing and quantum information theory \cite{Nielsen}, today being two of the most active fields of research in both theoretic and experimental physics. Among economists may he be mostly known as the father of modern game theory \cite{GT-Critical,GT-fudenberg,Course in GT}, the study of rational interactions in strategic situations. A field well rooted in the influential book\emph{ Theory of Games and Economic Behavior} (1944), by Von Neumann and Oscar Morgenstern. The book offered great advances in the analysis of strategic games and in the axiomatization of measurable utility theory, and drew the attention of economists and other social scientists to these subjects. For the last decade or so there has been an active interdisciplinary approach aiming to extend game theoretical analysis into the framework of quantum information theory, through the study of quantum games \cite{flitney,pitrowski1,NEinQ,landsburg,Bleiler,MW}; offering a variety of protocols where use of quantum peculiarities like entanglement in quantum superpositions, and interference effects due to quantum operations has shown to lead to advantages compared to strategies in a classical framework. The first papers appeared in 1999. Meyer showed with a model of a penny-flip game that a player making a \emph{quantum move} always comes out as a winner against a player making a \emph{classical} move regardless of the classical players choice \cite{meyer}. The same year Eisert et al. published a quantum protocol in which they overcame the dilemma in Prisoners dilemma \cite{eisert}. In 2003 Benjamin and Hayden generalized Eisert's protocol to handle multi-player quantum games and introduced the quantum minority game together with a solution for the four player case which outperformed the classical randomization strategy \cite{benjamin}. These results were later generalized to the $n$-players by Chen et al. in 2004 \cite{chen}. Multi-player minority games has since then been extensively investigated by Flitney et al. \cite{flitney1,flitney2,schmid}. An extension to multi-choice games, as the Kolkata resturant problem was offered by the authors of this review, in 2011 \cite{puya}.

\subsection{Games as information processing}
Information theory is largely formulated independent of the physical systems that contains and processes the information. We say that the theory is substrate independent. If you read this text on a computer screen, those bits of information now represented by pixels on your screen has traveled through the web encoded in electronic pulses through copper wires, as burst of photons trough fiber-optic cables and for all its worth maybe on a piece of paper attached to the leg of a highly motivated raven. What matters from an information theoretical perspective is the existence of a differentiation between some states of affairs. The general convention has been to keep things simple and the smallest piece of information is as we all know a \emph{bit} $b \in \{0,1\}$, corresponding to a binary choice: \emph{true} or \emph{false}, \emph{on} or \emph{off}, or simply \emph{zero} or \emph{one}. Any chunk of information can then be encoded in strings of bits: $\textbf{b}=b_{n-1} b_{n-2} \cdots b_0 \in \{0,1\}^n$. We can further define functions on strings of bits, $f: \{0,1\}^n \rightarrow \{0,1\}^k$ and call these functions computations or actions of information processing.

In a similar sense games are in their most general form independent of a physical realization. We can build up a formal structure for some strategic situation and model cooperative and competitive behavior within some constrained domain without regards to who or what these game playing agents are or what their actions actually is. No matter if we consider people, animals, cells, multinational companies or nations, simplified models of their interactions and the accompanied consequences can be formulated in a general form, within the framework of  game theory.

Lets connect these two concepts with an example. We can create a one to one correspondence with between the conceptual framework of game theory and the formal structure of information processing. Let there be $n$ agents faced with a binary choice of joining one of two teams. Each choice is represented by a binary bit $b_i \in \{0,1\}$. The final outcome of these individual choices is then given by a $n$-bit output string $\textbf{b} \in \{0,1\}^n$. We have $2^n$ possible outcomes, and for each agent we have some preference relation over these outcomes $\textbf{b}_{j}$. For instance, agent $1$ may prefer to have agent 3 in her team over agent 4, and may prefer any configuration where agent 5 is on the other team over any where they are on the same and so on. For each agent $i$, we'll have a preference relation of the following form, fully determining their objectives in the given situation:
\begin{equation}\label{pref}
\textbf{b}_{x_1}\succeq \textbf{b}_{x_2}\succeq \cdots \succeq \textbf{b}_{x_m}, \;\;\; m=2^n,
\end{equation}
where $\textbf{b}_{x_i}\succeq \textbf{b}_{x_j}$ means that the agent in question prefers $\textbf{b}_{x_i}$ to $\textbf{b}_{x_j}$, or is at least indifferent between the choices. To formalize things further we assign a numerical value to each outcome $\textbf{b}_{x_j}$ for \emph{each} agent, calling it the \emph{payoff} $\$_i(\textbf{b}_{x_j})$ to agent $i$ due to outcome $\textbf{b}_{x_j}$. This allows us to move from the preference relations in (\ref{pref}) to a sequence of inequalities. $\textbf{b}_{x_i}\succeq \textbf{b}_{x_j} \Longleftrightarrow \$(\textbf{b}_{x_i})\geq \$(\textbf{b}_{x_j})$. The aforementioned binary choice situation can now be formulated in terms of functions $\$_i(\textbf{b}_{x_j})$ of the output strings $\textbf{b}_{x_j}$, where each entry $b_i$ in the strings corresponds to the choice of an agent $i$. So far has the discussion only regarded the output string without mentioning any input. We could without loss of generality define an input as string where all the entries are initialized as 0's, and the individual choices being encoded by letting each participant either leave their bit unchanged or performing a NOT-operation, where $\textrm{NOT}(0)=1$.
More complicated situations with multiple choices could be modeled by letting each player control more than one bit or letting them manipulate strings of information bearing units with more states than two; of which we will se an example of later.

\subsection{Quantization of information}

Before moving on to the quantum formalism of operators and quantum states, there is one intermediate step worth mentioning, the \emph{probabilistic} bit, which has a certain probability $p$ of being in one state and a probability of $1-p$ of being in the other. If we represent the two states '0' and '1' of the ordinary bit by the two-dimensional vectors $(1,0)^T$ and $(0,1)^T$, then a probabilistic bit is  given by a linear combination of those basis vectors, with real positive coefficients $p_0$ and $p_1$, where $p_0+p_1=1$. In this formulation, randomization between two different choices in a strategic situation would translate to manipulating an appropriate probabilistic bit.

\paragraph{\textbf{The quantum bit}}

Taking things a step further, we introduce the quantum bit or the \emph{qubit}, which is a representation of a two level quantum state, such as the spin state of an electron or the polarization of a photon.
A qubit lives in a two dimensional complex space spanned by two basis states denoted $\left|0\right\rangle $
and $\left|1\right\rangle $, corresponding to the two states of the classical bit.

\begin{equation}
\left|0\right\rangle =\left(\begin{array}{c}
1\\
0
\end{array}\right),\;\left|1\right\rangle =\left(\begin{array}{c}
0\\
1
\end{array}\right).
\end{equation}
Unlike the classical bit, the qubit can be in any superposition of $\left|0\right\rangle $ and $\left|1\right\rangle $:
\begin{equation}
\left|\psi\right\rangle =a_{0}\left|0\right\rangle +a_{1}\left|1\right\rangle,
\end{equation}
where $a_{0}$ and $a_{1}$ are complex numbers obeying $|a_{0}|^{2}+|a_{1}|^{2}=1$. $|a_{i}^{2}|$ is simply the probability to find the system in the state
$\left|\,i\,\right\rangle ,\; i\in\{0,1\}$. Note the difference between this and the case of the probabilistic bit! We are now dealing with complex coefficients, which means that if we superpose two qubits, then some coefficients might be eliminated. This interference is one of many effects without counterpart in the classical case.  The state of an arbitrary qubit can be written in the \emph{computational basis} as:
\begin{equation}
\left|\psi\right\rangle =\left(\begin{array}{c}
a_{0}\\
a_{1}
\end{array}\right)
\end{equation}

The state of a general qubit can be parameterized as:

\begin{equation}
\left|\psi\right\rangle =\cos\left(\frac{\vartheta}{2}\right)\left|0\right\rangle +e^{i\varphi}\sin\left(\frac{\vartheta}{2}\right)\left|1\right\rangle,
\end{equation}

where we have factored out and omitted a global phase due to the physical equivalence between the states $e^{i\phi}\left|\psi\right\rangle$ and $\left|\psi\right\rangle$. This so called \emph{state vector} describes a point on a spherical surface with $\left|0\right\rangle $
and $\left|1\right\rangle $ at its poles, called the Bloch-sphere, parameterized by two real numbers $\theta$ and $\varphi$, depicted in figure 1.

\begin{figure}
\center
\includegraphics[scale=0.45]{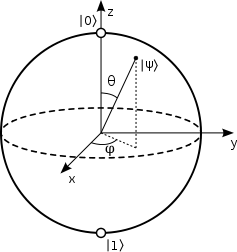}
\caption{Bloch sphere}
\end{figure}

\subsubsection{Hilbert spaces and composite systems}

The state vector of a quantum system is defined in a complex vector
space called \textit{Hilbert spac}e $\mathcal{H}$.
Quantum states are represented in common Dirac notation as ``ket's'',
written as the right part $\left|\psi\right\rangle $ of a bracket
(``bra-ket''). Algebraically a ``ket'' is column vector in our
state space. This leaves us to define the set of ``bra's'' $\langle\phi|$
on the dual space of $\mathcal{H}$, $\mathcal{H^{\star}}$. The dual Hilbert space $\mathcal{H^{\star}}$ is defined as the set of
linear maps $\mathcal{H}\rightarrow\textbf{C}$, given by
\begin{equation}
\langle\phi|\::\,\left|\psi\right\rangle \mapsto\langle\phi|\psi\rangle\in\mathbb{C},
\end{equation}
where $\langle\phi|\psi\rangle$ is the inner product of the vectors
$\left|\psi\right\rangle ,\left|\phi\right\rangle \in\mathcal{H}$.
We can now write down a more formal definition of a Hilbert
space: It is a complex inner product space with the following properties:
\begin{enumerate}
\item $\langle\phi|\psi\rangle=\langle\psi|\phi\rangle^{\dagger}$, where
$\langle\psi|\phi\rangle^{\dagger}$ is the complex conjugate of $\langle\psi|\phi\rangle$.
\item The inner product$ \langle\phi|\psi\rangle$ is linear in the first
argument: $\langle a\phi_{1}+b\phi_{2}|\psi\rangle=a^{\dagger}\langle\phi_{1}|\psi\rangle+b^{\dagger}\langle\phi_{2}|\psi\rangle$.
\item $\langle\psi|\psi\rangle\geq0$.
\end{enumerate}

The space of a $n$ qubit system is spanned by a basis of $2^{n}$ orthogonal
vectors $\left|e_{i}\right\rangle $; one for each  possible combination of the basis-states
of the individual qubits, obeying the orthogonality condition:
\begin{equation}
\langle e_{i}|e_{j}\rangle=\delta_{ij},
\end{equation}
where $\delta_{ij}=1$ for $i=j$ and $\delta_{ij}=0$ for $i\neq j$.
We say that the Hilbert space of a composite system is the tensor
products of the Hilbert spaces of its parts. So the space
of a $n$ qubit system is simply the tensor product of the spaces
of the $n$ qubits.
\begin{equation}
\mathcal{H}_{\mathcal{Q}}=\mathcal{H}_{\mathcal{Q}_n}\otimes\mathcal{H}_{\mathcal{Q}_n-1}\otimes\mathcal{H}_{\mathcal{Q}_{n-2}}...\otimes\mathcal{H}_{\mathcal{Q}_1},
\end{equation}
where $\mathcal{Q}_i$ the quantum system $i$ is a vector in $\textbf{C}^2$. A general $n$ qubit system can therefore be written
\begin{equation}\label{multi1}
\left|\psi\right\rangle =\sum_{x_{n},..,x_{1}=0}^{1}a_{x_{n}...x_{1}}\left|x_{n}\cdots x_{1}\right\rangle ,
\end{equation}
where
\begin{equation}\label{multi2}
\left|x_{n}\cdots x_{1}\right\rangle =\left|x_{n}\right\rangle \otimes\left|x_{n-1}\right\rangle \otimes\cdots\otimes\left|x_{1}\right\rangle \in\mathcal{H}_{\mathcal{Q}}
\end{equation}
with $x_{i}\in\{0,1\}$ and complex coefficients $a_{x_{i}}$.
For a two qubit system, $\left|x_{2}\right\rangle \otimes\left|x_{1}\right\rangle =\left|x_{2}\right\rangle \left|x_{1}\right\rangle =\left|x_{2}x_{1}\right\rangle $, we have
\begin{equation}
\left|\psi\right\rangle =\sum_{x_{2},x_{1}=0}^{1}a_{x_{2}x_{1}}\left|x_{2}x_{1}\right\rangle =a_{00}\left|00\right\rangle +a_{01}\left|01\right\rangle +a_{10}\left|10\right\rangle +a_{11}\left|11\right\rangle
\end{equation}
This state space is therefore spanned by four basis vectors:
\begin{equation}
\left|00\right\rangle ,\left|01\right\rangle ,\left|10\right\rangle ,\left|11\right\rangle ,
\end{equation}
which are represented by the following 4-dimensional column vectors
respectively:

\begin{equation}
\left(\begin{array}{c}
1\\
0\\
0\\
0
\end{array}\right),\left(\begin{array}{c}
0\\
1\\
0\\
0
\end{array}\right),\left(\begin{array}{c}
0\\
0\\
1\\
0
\end{array}\right),\left(\begin{array}{c}
0\\
0\\
0\\
1
\end{array}\right).
\end{equation}

\begin{figure}
\center
\includegraphics[scale=0.45]{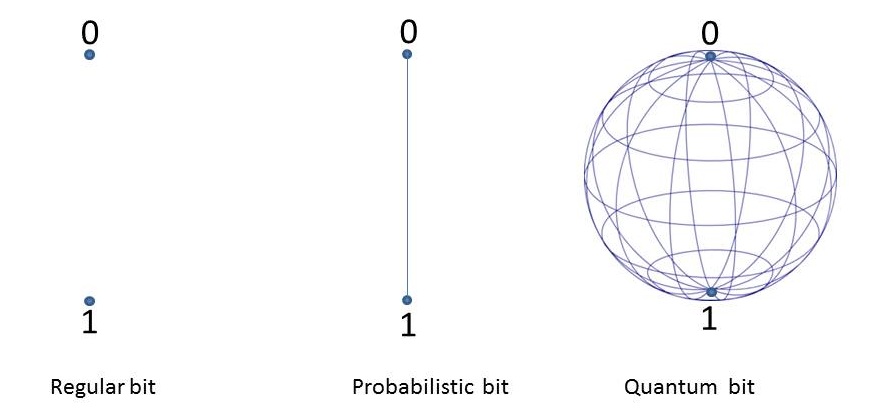}
\caption{The classical bit has only two distinct states, the probabilistic bit can be in any normalized convex combination of those states, whereas the quantum bit has a much richer state space.}
\end{figure}

\subsubsection{Operators}

A linear operator on a vector space $\mathcal{H}$ is a linear transformation
$\mathrm{T\,}:\mathcal{\: H}\rightarrow\mathcal{H}$, that maps vectors
in $\mathcal{H}$ to vectors in the same space $\mathcal{H}$. Quantum states
are normalized, and we wish to keep the normalization; we are therefore interested
in transformations that can be regarded as rotations in $\mathcal{H}$. Such transformations
are given by \textit{unitary operators} $U$. An operator $U$ is
called unitary if $U^{-1}=U^{\dagger}$. They preserve
inner products between vectors, and thereby their norm.
A \emph{projection operator} $P$ is Hermitian i.e. $P=P^{\dagger}$ and satisfies
$P^{2}=P$.
We can create a projector $P$, by taking the outer product
of a vector with itself:
\begin{equation}
P=\left|\phi\right\rangle \langle\phi|.
\end{equation}
$P$ is a matrix with every element $P_{ij}$ being the product of
the elements $i,j$ of the vectors in the outer product. This operator
projects any vector $\left|\gamma\right\rangle $ onto the 1-dimensional
subspace of $\mathcal{H}$, spanned by $\left|\phi\right\rangle $:

\begin{equation}
P\left|\gamma\right\rangle =\left|\phi\right\rangle \langle\phi|\left|\gamma\right\rangle =\langle\phi|\gamma\rangle\left|\phi\right\rangle .
\end{equation}
It simply gives the portion of $\left|\gamma\right\rangle $ along
$\left|\phi\right\rangle $.\\
We will often deal with unitary operators $U\in\mathrm{SU(2)}$, i.e
operators from the \textit{special unitary group} of dimension 2. The group
consists of $2\times2$ unitary matrices with determinant 1. These
matrices will be operating on single qubits (often in systems of 2
or more qubits). The generators of the group are the \textit{Pauli
spin matrices} $\sigma_{x},\,\sigma_{y},\,\sigma_{z}$, shown together with the identity
matrix $I$:
\begin{equation}
I=\left(\begin{array}{cc}
1 & 0\\
0 & 1
\end{array}\right),\;\sigma_{x}=\left(\begin{array}{cc}
0 & 1\\
1 & 0
\end{array}\right),\;\sigma_{y}=\left(\begin{array}{cc}
0 & -i\\
i & 0
\end{array}\right),\;\sigma_{z}=\left(\begin{array}{cc}
1 & 0\\
0 & -1
\end{array}\right).
\end{equation}
Note that $\sigma_x$ is identical to a classical (bit-flip) 'NOT'-operation.
General $2\times2$ unitary operators can be parameterized with three
parameters $\theta,\alpha,\beta$, as follows:
\begin{equation}\label{su2par}
U\left(\theta,\alpha,\beta\right)=\left(\begin{array}{cc}
e^{i\alpha}\cos\left(\theta/2\right) & ie^{i\beta}\sin\left(\theta/2\right)\\
ie^{-i\beta}\sin\left(\theta/2\right) & e^{-i\alpha}\cos\left(\theta/2\right)
\end{array}\right).
\end{equation}

An operation is said to be local if it only affects a part of a composite (multi-qubit) system. Connecting this to the concept of the bit-strings in the previous section; a local operation translates to just controlling one such bit. This is a crucial point in the case of modeling the effect of individual actions, since each agent in a strategic situation is naturally constrained to decisions regarding their own choices.
The action of a set of local operations on a composite system is given by the tensor product of the local operators. For a general n-qubit $\left|\psi\right\rangle$ as given in (\ref{multi1}) and (\ref{multi2}) we get:
\begin{equation}
U_{n}\otimes U_{n-1}\otimes \cdots \otimes U_{1}\left|\psi\right\rangle = \sum_{x_{n},..,x_{1}=0}^{1}a_{x_{n}...x_{1}}U_{n}\left|x_n\right\rangle\otimes U_{n-1}\left|x_{n-1}\right\rangle\otimes\cdots\otimes U_{1}\left|x_{1}\right\rangle.
\end{equation}

\subsubsection{Mixed states and the density operator}
We have so far only discussed \emph{pure states}, but sometimes we encounter quantum states without a definite state vector $\left|\psi\right\rangle $, these are called \emph{mixed states} and consists of a states that has certain probabilities of being in some
number of different pure states. So for example a state that is in $\left|\psi_{1}\right\rangle =a_{0}^{1}\left|0\right\rangle +a_{1}^{1}\left|1\right\rangle $ with probability $p_{1}$ and in $\left|\psi_{2}\right\rangle =a_{0}^{2}\left|0\right\rangle +a_{1}^{2}\left|1\right\rangle $ with probability $p_{2}$ is mixed. We handle mixed states by defining a density operator $\rho$, which is a hermitian matrix with unit trace:

\begin{equation}
\rho=\sum_{i}p_{i}\left|\psi_{i}\right\rangle \langle\psi_{i}|,
\end{equation}
where $\sum_i p_i = 1$. A pure state in this representation is simply a state for which all probabilities, except one is zero.
If we apply a unitary operator $U$ on a pure state, we end up with $U\left|\psi\right\rangle $ which has the density
operator $U\rho U^{\dagger}=U\left|\psi\right\rangle \langle\psi|U^{\dagger}$.
Regardless if we are dealing with pure or mixed states, we take the expectation value of upon measurement
ending up in a $\left|\phi\right\rangle $
by calculating $\mathrm{Tr}\left(\left|\phi\right\rangle \langle \phi|\rho\right)$, where $|\phi\rangle \langle \phi|$ is a so called projector. For calculating the expectation values of a state to be in \emph{any} of a number of states $|\phi_i\rangle$, we construct a projection operator $P=\sum_i |\phi_i\rangle \langle \phi_i|$ and take the trace over $P$ multiplied by $\rho$.

\subsubsection{Entanglement}
Entanglement is the resource our game-playing agents will make use of in the quantum game protocols to achieve better than classical performance. Non-classical correlations are thus introduced, by which the players can synchronize their behavior without any additional communication. An entangled state is basically a quantum system that \emph{cannot} be written as a tensor product of its subsystems, we'll thus define two classes of quantum states. Examples below refers to two-qubit states.

Product states:
\begin{equation}|\Psi_{\mathcal{Q}}\rangle=|\Psi_{\mathcal{Q}_{2}}\rangle\otimes|\Psi_{\mathcal{Q}_{1}}\rangle, ~~\textrm{or using density matrix} ~~\rho_{\mathcal{Q}}=\rho_{\mathcal{Q}_{2}}\otimes
\rho_{\mathcal{Q}_{1}},
\end{equation}

and entangled states
\begin{equation}|\Psi_{\mathcal{Q}}\rangle\neq\ket{\Psi_{\mathcal{Q}_{2}}}\otimes|\Psi_{\mathcal{Q}_{1}}\rangle, ~~\textrm{or using density matrix }~~\rho_{\mathcal{Q}}\neq\rho_{\mathcal{Q}_{2}}\otimes
\rho_{\mathcal{Q}_{1}}.
\end{equation}

For a mixed state, the density matrix is defined as mentioned by
$\rho_{\mathcal{Q}}=\sum^N_{i=1}p_{i}|\psi_{i}\rangle\langle\psi_{i}|$ and it is said to be separable, which we will denote
by $\rho^{sep}_{\mathcal{Q}}$, if it can  be written as
\begin{equation}\label{eq:sep}
\rho^{sep}_{\mathcal{Q}}=\sum_{i}p_{i}(
\rho^i_{\mathcal{Q}_{2}}\otimes \rho^i_{\mathcal{Q}_{1}}),~\sum_{i}p_{i}=1.
\end{equation}

A set of very important two-qubit entangled states  are the Bell states
\begin{equation}|\Phi^{\pm}_{\mathcal{Q}}\rangle=
 \frac{1}{\sqrt{2}}(|{00}\rangle\pm|11\rangle),~~~~|\Psi^{\pm}_{\mathcal{Q}}\rangle=
 \frac{1}{\sqrt{2}}(|{01}\rangle\pm|10\rangle).
\end{equation}

The GHZ-type-states
\begin{equation}
|\textrm{GHZ}_n\rangle=\frac{1}{\sqrt{2}}\left ( |00\cdots 0\rangle + e^{i\phi}|11\cdots 1\rangle \right)
\end{equation}
could be seen as a $n$-qubit generalization of $|\Phi^{\pm}_{\mathcal{Q}}\rangle$-states.

\subsection{Classical Games}
\label{sec:1}
It is instructive to review the theory of classical games and some major solution concepts before moving on to examples of quantum games.
We'll start by defining classical pure and mixed strategy games, and then move on to introducing some relevant solution concepts and finish off with a definition of quantum games.

A game is a formal model over the interactions between a number of agents (\emph{agents, players, participants}, and \emph{decision makers} may be used interchangeably) under some specified sets of choices (\emph{choices, strategies, actions} and \emph{moves}, may be used interchangeably). Each combination of choices made, or strategies chosen by the different players leads to an outcome with some certain level of desirability for each of them. The level of desirability is measured by assigning a real number, a so called \emph{payoff} $\$$ for each game outcome for each player. Assuming rational players, each will choose actions that maximizes their expected payoff $E(\$)$, i.e. in an deterministic as well as in an probabilistic setting acting in a way that, based on the known information about the situation, maximizes the expectation value of their payoff.
The structure of the game is fully specified by the relations between the different combinations of strategies and the payoffs received by the players. A key point is the interdependence of the payoffs with the strategies chosen by the other players. A situation where the payoff of one player is independent of the strategies of the others would be of little interest from a game theoretical point of view.  It is natural to extend the notion of payoffs to \emph{payoff functions} whose arguments are the chosen strategies of all players and ranges are the real valued outputs that assigns a level of desirability for each player to each outcome.

\paragraph{\textbf{Pure strategy classical game}}
We have a set of $n$ players $\{1,2,...,n\}$, $n$ strategy sets $S_i$, one for each player $i$, with $s_i^j \in S_i$, where $s_i^j$ is the $j$:th strategy of player $i$. The strategy space $S=S_1\times S_2\times\cdots \times S_n$ contains all $n$-tuples pure strategies, one from each set. The elements $\sigma \in S$ are called strategy profiles, some of which will earn them the status of being a \emph{solution} with regards to some solution concept.

We define a game by its payoff-functions $\$_i$, where each is a mapping from the strategy space $S$ to a real number, the payoff or utility of player $i$. We have:

\begin{equation}
\$_{i}: S_1\times S_2\times\cdots \times S_n \rightarrow \textbf{R}.
\end{equation}

\paragraph{\textbf{Mixed strategy classical game}}
Let $\Delta(S_i)$ be the set of convex linear combinations of the elements $s_i^j \in S_i$. A mixed strategy $s^{mix}_i \in \Delta(S_i)$ is then given by:

\begin{equation}
\sum_{s^{j}_i \in S_i} p_i^j s^j_i \;\;\; \textrm{with} \;\;\;  \sum_{j} p_i^j = 1,
\end{equation}

where $p^j_i$ is the probability player $i$ assigns to the choice $s^j_i$.  The space of mixed strategies $\Delta(S)=\Delta(S_1)\times \Delta(S_2)\times \cdots \times \Delta(S_n)$ contains all possible mixed strategy profiles $\sigma_{mix}$. We now have:

\begin{equation}
\$_i: \Delta(S_1)\times \Delta(S_2)\times \cdots \times \Delta(S_n)\rightarrow \textbf{R}.
\end{equation}

Note that the pure strategy games are fully confined within the definition of mixed strategy games and can be accessed by assigning all strategies except one, the probability $p^j = 0$. This class of games could be formalized in a framework using probabilistic information units, such as the probabilistic bit.

\subsection{\textbf{Solution concepts}}
We will introduce two of many game theoretical solution concepts. A solution concept is a strategy profile $\sigma^* \in S$, that has some particular properties of strategic interest. It could be a strategy profile that one would expect a group of rational self-maximizing agents to arrive at in their attempt to maximize their minimum expected payoff. Strategy profiles of this form i.e. those that leads to a combination of choices where each choice is the best possible response to any possible choice made by other players tend to lead to an equilibrium, and are good predictors of game outcomes in strategic situations. To see how such equilibria can occur we'll need to develop the concept of \emph{dominant strategies}.

\definition{(\textbf{Strategic dominance}):}
A strategy $s^{dom}\in S_i$ is said to be dominant for player $i$, if for any strategy profile $\sigma_{-i} \in S/S_i$, and any other strategy $s^j\neq s^{dom} \in S_i$:
\begin{equation}
  \$_i(s^{dom},\sigma_{-i})\geq \$_i(s^{j},\sigma_{-i}) \;\;\textrm{for all}\;\; i=1,2, \cdots ,n.
\end{equation}

Lets look at a simple example. Say that we have two players, Alice with legal strategies $s^1_{Alice}, s^2_{Alice} \in S_{Alice}$ and Bob with $s^1_{Bob}, s^2_{Bob} \in S_{Bob}$. Now, if the payoff Alice receives when playing $s^1_{Alice}$ against any of Bob's two strategies is higher than (or at least as high as) what she receives by playing $s^2_{Alice}$, then $s^1_{Alice}$ is her dominant strategy. Her payoff can of course vary depending on Bob's move but regardless what Bob does, her dominant strategy is the \emph{best response}. Now there is no guarantee that such dominant strategy exists in a pure strategy game, and often must the strategy space be expanded to accommodate for mixed strategies for them to exist.

If both Alice and Bob has a dominant strategy, then this strategy profile becomes a \emph{Nash Equilibrium}, i.e. a combination of strategies for which none of them can gain by unilaterally deviating from. The Nash equilibrium profile acts as an attractor in the strategy space and forces the players into it, even though it is not always an optimal solution. Combinations can exist that can lead to better outcomes for both (all) players.

\definition{(\textbf{Nash equilibrium}):}
Let $\sigma^{NE}_{-i} \in S/S_i $ be a strategy profile containing the dominant strategies of every player except player $i$, and let $s^{NE}_i\in S_i$ be the the dominant strategy of player $i$. Then for all $s^j_i \neq s^{NE}_i \in S_i$:
\begin{equation}
\$_i(s^{NE}_i,\sigma^{NE}_{-i})\geq \$_i(s^{j}_i,\sigma^{NE}_{-i}) \;\;\textrm{for all}\;\; i=1,2, \cdots ,n.
\end{equation}

If we have a situation where an agent can increase its payoff without decreasing any others, then this would per definition mean that nobody would mind if that agent would do so. Each such increase in payoff is called a \emph{Pareto improvement}. When no such improvement can be done, then the strategy profile is said to be \emph{Pareto optimal}.

\definition{(\textbf{Pareto efficiency}):}
A Pareto efficient or Pareto optimal strategy profile is one where none of the participating agents can increase their payoff without decreasing the payoff of someone else.

\section{Quantum Games}

In the quantum game protocols (\emph{protocol} and \emph{scheme} may be used interchangeably) presented in this paper, the $m_i$ different choices available to a player $i$ will be encoded in the basis states of an $m_i$-level quantum system, where the $m_i$ denotes the dimensionality of the Hilbert space $\mathcal{H}_{\mathcal{Q}_i}$ associated with that subsystem. Each of the $n$ player holds one subsystem leading to a total system with a state vector a in an $\prod_{i=1}^n  \textrm{dim}(\mathcal{H}_{\mathcal{Q}_i})$ - dimensional space. The definition of a quantum game must therefore include a Hilbert space of a multipartite multilevel system $\mathcal{H}_{\mathcal{Q}}=\mathcal{H}_{\mathcal{Q}_n}\otimes \mathcal{H}_{\mathcal{Q}_{n-1}}\cdots \otimes \mathcal{H}_{\mathcal{Q}_1}$.

The different subsystems must in general be allowed to have a have a common origin to accommodate entanglement in the shared initial state $\rho_{in}\in \mathcal{H}_{\mathcal{Q}}$. This is often modeled by including a referee that prepares an initial state and distributes the subsystems among the players. Wether or not this step invokes on the non-communication criteria certain games have, is under debate. We justify it by the fact that no communication is done under the crucial step of choosing a strategy. The strategies are applied by local quantum operations on the quantum state held by each player. No player has any access to any part of the system except its own subsystem, and no information can be sent between the players with aid of the shared quantum resource.
Classical strategies becomes quantum strategies by expanding the strategy sets:

\begin{equation}
s_i \in S_i \; \Rightarrow \;U_i \in \textrm{S}(m_i),
\end{equation}

where the set of allowed quantum operations $\textrm{S}(m_i)$ is some subset of the special unitary group $\textrm{SU}(m_i)$. We will later see that the nature of the game can be determined by restrictions on $\textrm{S}(m_i)$. It is an important point to be able to show that the classical version of a game is recoverable just by restricting the set of allowed operators. At least if we want it to be a \emph{proper quantization} \cite{Bleiler}, i.e. an extension of the classical game into the quantum realm, and not a whole new game without a classical counterpart.

We define a quantum game in two steps:

\begin{equation}
 \; U_n\otimes U_{n-1} \otimes \cdots \otimes U_1:\mathcal{H}_{\mathcal{Q}_{n}}\otimes \mathcal{H}_{\mathcal{Q}_{n-1}}\cdots \otimes \mathcal{H}_{\mathcal{Q}_1} \rightarrow \mathcal{H}_{\mathcal{Q}_{n}}\otimes\mathcal{H}_{\mathcal{Q}_{n-1}}\cdots \otimes \mathcal{H}_{\mathcal{Q}_1},
\end{equation}

\begin{equation}
 \; \$_i: \mathcal{H}_{\mathcal{Q}_{n}}\otimes \mathcal{H}_{\mathcal{Q}_{n-1}}\cdots \otimes \mathcal{H}_{\mathcal{Q}_1} \rightarrow \textbf{R},
\end{equation}

where the first step is a transformation of the state of the complete system by local operations, and the second is a mapping from the Hilbert space of the quantum state to a real number, the expected payoff of player $i$.

\subsection{The quantum game protocol}

\begin{itemize}
  \item The game begins with an entangled initial state $|\psi_{in}\rangle$. Each subsystem has a dimensionality $m$ that equal to the number of pure strategies in each players strategy set. In the protocols covered in this paper, all players will face the same number of choices. The number of subsystems equals the number of players. One can assume that $|\psi_{in}\rangle$ has been prepared at some location by a referee that then has distributed the subsystems among the players \cite{eisert,benjamin}.

  \item The players then chooses an unitary operator $U$ from a subset of SU($m$), and applies it to their subsystem. The initial state $\rho_{in}$ transforms to a final state $\rho_{fin}$, given by:

      \begin{equation}
       \rho_{fin}=U\otimes U\otimes \cdots \otimes U \rho_{in} U^{\dagger}\otimes U^{\dagger}\otimes \cdots \otimes U^{\dagger}
      \end{equation}
      In the absence of communication, and due to the symmetry of these games, all players are expected to do the same operation.

  \item The players then measures their own subsystem, collapsing their quantum states to units of classical information. For the case of a two-choice protocol, each player ends up with a classical bit $b_i$, and the complete system has thus collapsed into a classical string $\textbf{b}$, corresponding to a pure strategy profile $\sigma \in S$. For the quantum game to have an advantage over a classical game, the collective action of the players must have decreased the probability of the final state $\rho_{fin}$ to collapse into such basis states (classical information strings / strategy profiles) that are undesired, i.e. leading to lower or zero payoff $\$$.

  \item  To calculate the expected payoffs $E(\$)$, we define for each player $i$  a payoff-operator $P_{i}$ , which contains the sum of orthogonal projectors associated with the states for which player $i$  receives a payoff $\$$. We have:
  \begin{equation}\label{payoffoperator}
      P_i = \sum_j \$^j_i|\chi^{j}_i\rangle \langle\chi^{j}_i|,
      \end{equation}
      where the states $|\chi^{j}_i\rangle$ are those sates that leads to a payoff for player $i$, and $\$^j_i$ the associated payoffs. The expected payoff $E(\$_{i})$  of player $i$  is calculated by taking the trace of the product of the final state $\rho_{fin}$  and the payoff-operator $P_{i}$:
\begin{equation}
E_{i}(\$)=\mathrm{Tr\left(\mathit{P}_{i}\rho_{fin}\right)}.
\end{equation}

\end{itemize}

\subsection{Prisoners dilemma}
The prisoners dilemma is one of the most studied game theoretical problems. It was introduced in 1950 by Merrill Flood and Melvin Dresher, and has been widely used ever since to model a variety of situations, including oligopoly pricing, auction bidding, salesman effort, political bargaining and arms races. In is in its standard form, a symmetric simultaneous game of complete information. Two players, Alice and Bob (A and B) are faced with a choice to \emph{cooperate} or to \emph{defect}, without any information about the action taken by the other. The payoffs they receive due to any combination of choices can be read of the table below, where the first entry in each parenthesis shows the payoff $\$_A$ of Alice and the second entry the payoff $\$_B$ of Bob.
\begin{table}\label{payoffmatrix}

\centering
\begin{tabular}{c|c|c|c|}
\multicolumn{1}{c}{} & \multicolumn{1}{c}{} & \multicolumn{2}{c}{Bob}\\
\cline{3-4}
\multicolumn{1}{c}{} &  & Cooperate & Defect\\
\cline{2-4}
\multirow{2}{*}{Alice  } & Cooperate & (3,3) & (0,5)\\
\cline{2-4}
 & Defect & (5,0) & (1,1)\\
\cline{2-4}
\end{tabular}
 \caption{The normal-form representation of prisoners dilemma.}\label{PDmatrix}
\end{table}
Given that Bob chooses to cooperate, Alice receives $\$_A=3$ if she chooses to do the same, and she receives $\$_A=5$ if she chooses to defect. If Bob instead defects, then Alice receives $\$_A=0$ by cooperating and $\$_A=1$ by choosing to defect. No matter what Bob does, Alice will always gain by choosing to defect, equipping her with a strictly dominant strategy! Due to the symmetry of the game, the same is true for Bob, forcing them into a Nash equilibrium strategy profile of (defect, defect), which pays out $\$_{AB}=1$ to each. This outcome is clearly far from efficient, since there is a Pareto optimal strategy profile (cooperate, cooperate) that would have given them $\$_{AB}=3$, and hence the dilemma.

Quantum prisoners dilemma was introduced by J. Eisert, M. Wilkens, and M. Lewenstein in 1999 \cite{meyer}. Here Alice and Bob are equipped with a quantum resource, a maximally entangled Bell-type-state, and each of them are in possession of a subsystem. The Hilbert space of the game is given by: $\mathcal{H}=\mathcal{H}_{B}\otimes\mathcal{H}_{A}$, with $\mathcal{H}_{A}=\mathcal{H}_{B}=\textbf{C}^{2}$. We'll identify the following relations, mapping classical outcomes with basis states of the Hilbert space: $(\mathrm{cooperate,cooperate})\rightarrow\left|00\right\rangle$, $(\mathrm{cooperate,defect})\rightarrow\left|01\right\rangle$, $(\mathrm{defect,cooperate})\rightarrow\left|10\right\rangle$ and $(\mathrm{defect,cooperate})\rightarrow\left|11\right\rangle$. The entangled initial state is created by acting with an entangling operator $J=\frac{1}{\sqrt{2}}I^{\otimes 2}+i\sigma_x^{\otimes 2}$ on a product state initialized as (cooperate, cooperate):
\begin{equation}
J\left|00\right\rangle=\frac{1}{\sqrt{2}}(\left|00\right\rangle +i\left|11\right\rangle).
\end{equation}
Note that the entangling operator performs a global operation, i.e. an operation performed on both subsystems simultaneously. One can consider it to be performed by a a referee, loyal to both parties. The game proceeds by Alice and Bob performing their local strategies $U_A$ and $U_B$, and the state is turned into its final form: $|\psi_{fin}\rangle=(U_B\otimes U_A) J|00\rangle$. Before measurement is performed, an disentangling operator $J^{\dagger}$ is applied. The inclusion of $J$ and $J^{\dagger}$ into the protocol assures that the classical game is embedded into the quantum version, whereby the classical prisoners dilemma can be accessed by restricting the set of allowed operators to $U_A, U_B \in \{I, \sigma_x\}$. It is a simple task to show that any combination of the identity operator $I$ and the bit-flip operator $\sigma_x$ commutes with $J$, and together with the fact that $JJ^{\dagger}=I$, one concludes that this restriction turns the protocol into classical (one-bit) operations on a bit string '00'.

\begin{figure}
\center
\includegraphics[scale=0.50]{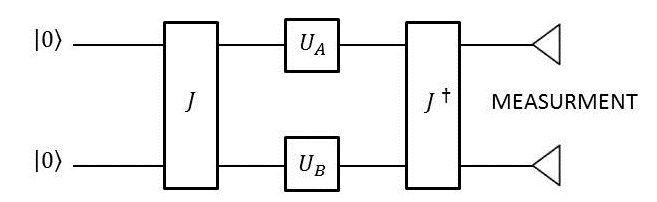}
\caption{Circuit diagram of the quantum prisoners dilemma protocol.}
\end{figure}

It is now left to define a set of operators $U$, representing allowed \emph{quantum} strategies, and the payoff operators $P_A$ and $P_B$. Eisert et.al. considered a two parameter subset of SU(2) as the strategy space:

\begin{equation}
U\left(\theta,\alpha\right)=\left(\begin{array}{cc}
e^{i\alpha}\cos\left(\theta/2\right) & \sin\left(\theta/2\right)\\
-\sin\left(\theta/2\right) & e^{-i\alpha}\cos\left(\theta/2\right)
\end{array}\right).
\end{equation}

The classical strategies are represented by $U(0,0)=I$ and $U(0,\pi)=\sigma_x$.  We construct Alice's payoff operator $P_A$ as defined in (\ref{payoffoperator}) with values from the payoff matrix:
\begin{equation}
P_A=3|00\rangle \langle 00|+5|01\rangle \langle 01|+1|11\rangle \langle 11|.
\end{equation}

Her expected payoff is calculated by taking the trace of the final state and the payoff operator: $E(\$_A)=\textrm{Tr}(P_A\rho_{fin})$, where $\rho_{fin}=|\psi_{fin}\rangle \langle\psi_{fin}|$.
It can be shown that when the set of strategies are expanded to allow any $U(\theta,\alpha)$, the old Nash equilibrium (defect, defect) $\rightarrow U(0,\pi)\otimes U(0,\pi)$ ceases to exist! Instead a new Nash equilibrium emerges at

\begin{equation}
U_A=U_B=U(0,\pi/2)= \left(\begin{array}{cc}
i & 0\\
0 & -i
\end{array}\right).
\end{equation}

This strategy leads to an expected payoff $E(\$_A)=E(\$_A)=3$. Thereby they both receive an expected payoff that equals the Pareto optimal solution in the classical pure strategy version, with the addition that this solution is also a Nash equilibrium. Dilemma resolved. It should be added that if the strategy sets are further expanded to include all SU(2) operations, this solution vanshes, and there is no Nash equilibrium strategy profile in pure quantum strategies, whereby one has to include mixed quantum operations to find an equilibrium \cite{comment}.

\subsection{Minority games}
 We extend the previous protocol to ones with multiple agents, by introducing the minority game. The game consists of $n$ of non-communicating players that must independently make up their mind between two choices. We could regard these players as investors on a market deciding between two equally attractive securities, as commuters choosing between two equally fast routes to a suburb, or any collection of agents facing situations where they wish to make the minority choice. The core objective of the players are thus to avoid the crowd. We encode the two choices as $\left|0\right\rangle $ and $\left|1\right\rangle $ in the computational basis like before. The players receive payoff a $\$=1$ if they happen to be in the smaller group. So if the number of players choosing $\left|0\right\rangle $ is less than the number of players choosing $\left|1\right\rangle $, the first group receives payoff whereas the second group is left with nothing. Would the players happen to be evenly distributed between the two choices, then they'll all go empty handed.

The Nash equilibrium solution is to randomize between $\left|0\right\rangle $ and $\left|1\right\rangle $ using a fair coin. The \emph{one shoot} version we are considering will necessarily have a mixed strategy solution, since any deterministic strategy would lead all players to the same choice and thus a maximally undesired outcome.  The expected payoff $E(\$)$ for a player is simply the number outcomes with that player in the minority group divided by the number of different possible outcomes. For a four player game, there are two minority outcomes for each player, out of sixteen possible. This gives a expected payoff of $1/8.$

A quantum version of a four player minority game was presented by Benjamin and Hayden in 2000 \cite{benjamin}, offering a solution that significantly outperformed the classical version of the game. The advantage comes from the possibility of eliminating (or reducing the probability of) such final outcomes where the players are evenly distributed among the two choices. The collective application of local unitary operators on the subsystems of an entangled state can thus transform this initial state in such a way that a better-than-classical result is achieved. This transformation does not have a classical analogue, and the performance is due to interference effects from the local phases added to the qubits by the players local operations.
We are not including the action of an entangling operator $J$ in this section, we simply assume the initial state to be entangled at the start of the protocol, and it can again be assumed that the state has been prepared by an unbiased referee and distributed among the players. Considering the four-player case, we begin the protocol with an GHZ-type state similar to the one used in the previous two-player game, but now consisting of \emph{four} entangled qubits.
\begin{equation}
\left|\psi_{in}\right\rangle=\frac{1}{\sqrt{2}}(\left|0000\right\rangle +\left|1111\right\rangle).
\end{equation}
The Hilbert space of the game is sixteen dimensional, accounting for all possible game outcomes. $\mathcal{H}_{\mathcal{Q}}=\mathcal{H}_{\mathcal{Q}_{4}}\otimes\mathcal{H}_{\mathcal{Q}_{3}}\otimes\mathcal{H}_{\mathcal{Q}_{2}}\otimes\mathcal{H}_{\mathcal{Q}_{1}}$, with $\mathcal{H}_{\mathcal{Q}_{i}}=\textbf{C}^2$. Each player $i=1,2,3,4$ is permitted to manipulate its subsystem with the full machinery of local quantum operations: $U_i \in \textrm{SU(2)}$ given in (\ref{su2par}).
The payoff operator $P_i$ projects the final state onto the desired states of player $i$, and is given by
\begin{equation}
P_{i}=\sum_{j=1}^{k}|\xi_{i}^{j}\rangle \langle \xi_{i}^{j}|.
\end{equation}
The sum is over all the $k$ different states $|\xi_{i}^{j}\rangle $,
for which player $i$ is in the minority. Its worth to note that the sums are always over a even number $k$, and that they run over the
states of the following form:
\begin{equation}
P_{i}=\sum_{j=1}^{k}|\xi_{i}^{j}\rangle \langle \xi_{i}^{j}|=\sum_{j=1}^{k/2}|\vartheta_{i}^{j}\rangle \langle \vartheta_{i}^{j}|+\sum_{j=1}^{k/2}|\overline{\vartheta_{i}^{j}}\rangle \langle \overline{\vartheta_{i}^{j}}|,
\end{equation}
where $|\overline{\vartheta_{i}^{j}}\rangle $ is the bit-flipped version of $|\vartheta_{i}^{j}\rangle $, i.e 0's and 1's
are interchanged. The payoff operator $P_{1}$ for player 1 in the four player case is given by:
\begin{equation}
P_{1}=\left|0001\right\rangle \left\langle 0001\right|+\left|1110\right\rangle \left\langle 1110\right|.
\end{equation}
By playing $U\left(\theta,\alpha,\beta\right)=U(\frac{\pi}{2},-\frac{\pi}{8},\frac{\pi}{8})$, the four players can completely eliminate the risk of upon measurement ending up with an outcome where none of them receives a payoff. This quantum strategy leads to an expected payoff $E(\$)=\frac{1}{4}$ that is twice as good as in the classical case $E(\$)=\frac{1}{8}$. The strategy profile is a Nash equilibrium as well as Pareto optimal. Quantum minority games has been extensively studied for cases of arbitrary $n$, and it can be shown that the quantum versions gives rise to better than classical payoffs for any game with an even number of players \cite{chen}.

\subsection{Kolkata restaurant problem}
The Kolkata restaurant problem is an extension of the minority game \cite{kolkata,kolkata1,kolkata2,kolkata3,Elfarol}, where the $n$ players now has $m$ choices. As the story goes, the choice is between $m$ restaurants. The players receive a payoff if their choice is not too crowded, i.e the number of agents that chose the same restaurant is under some limit. We will discuss the case for which this limit is one. Just like in the minority game previously discussed, the Kolkata restaurant problem offers a way for modeling heard behavior and market dynamics, where visiting a restaurant translates to buying a security, in which case an agent wishes to be the only bidder. In our simplified model there are just three agents, Alice, Bob and Charlie. They have three possible choices: security 0, security 1 and security 2. They receive a payoff $\$=1$ if their choice is unique, i.e that nobody else has made the same choice, otherwise they receive $\$=0$. The game is so called \emph{one shoot}, which means that it is non-iterative, and the agents have no information from previous rounds to base their decisions on. Under the constraint that they cannot communicate, there is nothing left to do other than randomizing between the choices just like in the minority games in the previous section. Given the symmetric nature of the problem, any deterministic strategy would lead all three agents to the same strategy, which in turn would mean that all three would leave empty handed. There are $27$ different strategy profiles possible, i.e combinations of choices. $12$ of which gives a payoff of $\$=1$ to each one of them. Randomization gives therefore agent $i$ an expected payoff of $E(\$)=\frac{4}{9}$.

In the quantum version we let Alice, Bob and Charlie share a quantum resource \cite{puya}. Each has a part of a multipartite quantum state. They play their strategy by manipulating their own part of the combined system, before measuring their subsystems and choosing accordingly. Whereas classically the players would be allowed randomizing over a discrete set of choices,  in the quantum version each subsystem is allowed to be transformed with arbitrary local quantum operations, just like before.
In the absence of entanglement, quantum games of this type usually yield the same payoffs as their classical counterparts, whereas the combination of unitary operators (or a subset therein) and entanglement, will be shown to outperform the classical randomization strategy.

When moving from quantum game protocols with two choices into ones with three, we'll need some additional structure. Instead of qubits will we be dealing with qutrits, which are their three level versions. The local operations on qutrits are now represented by a more complicated group of matrices, the SU(3) group. Everything else will essentially be similar to that of the quantum minority game.

A qutrit is a 3-level quantum system on 3-dimensional Hilbert space $\mathcal{H_\mathcal{Q}}=\textbf{C}^{3}$ , written in the computational basis as:

\begin{equation}
  |\psi\rangle=a_{0}|0\rangle+a_{1}|1\rangle+a_{2}|2\rangle \in \textbf{C}^3,
\end{equation}
with $a_{0},a_{1},a_{2}\in\textbf{C}$  and $|a_{0}|^{2}+|a_{1}|^{2}+|a_{2}|^{2}=1 $. A general $n$-qutrit system $\left|\Psi\right\rangle$   is a vector on $3^{n}$-dimensional Hilbert space, and is written as a linear combination of $3^{n}$ orthonormal basis vectors.

\begin{equation}
  \left|\Psi\right\rangle =\sum_{x_{n},..,x_{1}=0}^{2}a_{x_{n}...x_{1}}\left|x_{n}\cdots x_{1}\right\rangle ,
\end{equation}
where
\begin{equation}
\left|x_{n}\cdots x_{1}\right\rangle =\left|x_{n}\right\rangle \otimes\left|x_{n-1}\right\rangle \otimes\cdots\otimes\left|x_{1}\right\rangle \in\mathcal{H_\mathcal{Q}}=\overbrace{\textbf{C}^{3}\otimes...\otimes\textbf{C}^{3}}^\textrm{$n$-times},
\end{equation}
with $x_{i}\in\{0,1,2\}$  and complex coefficients $a_{x_{i}}$,  obeying $\sum|a_{x_{n}...x_{1}}|^{2}=1$.

Single qutrits are transformed with unitary operators $U \in$ SU(3), i.e operators from the special unitary group of dimension 3, acting on $\mathcal{H_\mathcal{Q}}$ as $U:\mathcal{H_\mathcal{Q}} \rightarrow \mathcal{H_\mathcal{Q}}$. In a multi-qutrit system, operations on single qutrits are said to be local. They affect the state-space of the corresponding qutrit only. The SU(3) matrix is parameterized by defining three general, mutually orthogonal complex unit vectors $\bar{x}, \bar{y}, \bar{z}$, such that $\bar{x}\cdot\bar{y}=0$  and $\bar{x}^*\times\bar{y}=\bar{z}$. We construct a SU(3) matrix by placing $\bar{x}, \bar{y}^*$ and $\bar{z}$  as its columns \cite{coherent}.
Now a general complex unit vector is given by:

\begin{equation}
\bar{x}=\left(\begin{array}{c}
\sin\theta\cos\phi e^{i\alpha_{1}}\\
\sin\theta\sin\phi e^{i\alpha_{2}}\\
\cos\theta e^{i\alpha_{3}}
\end{array}\right),
\end{equation}
and one complex unit vector orthogonal to $\bar{x}$  is given by:

\begin{equation}
\bar{y}=\left(\begin{array}{c}
\cos\chi\cos\theta\cos\phi e^{i(\beta_{1}-\alpha_{1})}+\sin\chi\sin\phi e^{i(\beta_{2}-\alpha_{1})}\\
\cos\chi\cos\theta\sin\phi e^{i(\beta_{1}-\alpha_{2})}-\sin\chi\cos\phi e^{i(\beta_{2}-\alpha_{2})}\\
-\cos\chi\sin\theta e^{i(\beta_{1}-\alpha_{3})}
\end{array}\right),
\end{equation}
where $0\leq\phi,\theta,\chi,\leq\pi/2$ and $0\leq\alpha_1,\alpha_2,\alpha_3,\beta_1,\beta_2\leq2\pi$. We have a general SU(3) matrix $U$, given by:

\begin{equation}
U=\left(\begin{array}{ccc}
x_{1} & y_{1}^{*} & x_{2}^{*}y_{3}-y_{3}^{*}x_{2}\\
x_{2} & y_{2}^{*} & x_{3}^{*}y_{1}-y_{1}^{*}x_{3}\\
x_{3} & y_{3}^{*} & x_{1}^{*}y_{2}-y_{2}^{*}x_{1}
\end{array}\right),
\end{equation}
and it is controlled by eight real parameters ${\phi,\theta,\chi,\alpha_1,\alpha_2,\alpha_3,\beta_1,\beta_2}$.

The initial state, a maximally entangled GHZ-type state
\begin{equation}\label{eq:GHZ}
\mid\psi_{in}\rangle=\frac{1}{\sqrt{3}}\left(|000\rangle+|111\rangle+|222\rangle\right) \in \mathcal{H_\mathcal{Q}}=\mathbb{C}^{3}\otimes\mathbb{C}^{3}\otimes\mathbb{C}^{3},
\end{equation}
 is symmetric and unbiased in regards to permutation of player position and has the property of letting us embed the classical version of the game, accessible trough restrictions on the strategy sets. To show this, we define a set of operators corresponding to classical pure strategies that gives raise to deterministic payoffs when applied to $\mid\psi_{in}\rangle$. The cyclic group of order three, $C_3$, generated by the matrix:

 \begin{equation}
 s=\ \left( \begin{array}{ccc}
0 & 0 & 1 \\
1 & 0 & 0 \\
0 & 1 & 0 \end{array} \right)\,
 \end{equation}
where $s^3=s^0=I$ and $s^2=s^T$, has the properties we are after. The set of classical strategies $S=\{s^0,s^1,s^2\}$ with $ s^i\otimes s^j \otimes s^k|000\rangle=|i\, j\, k\rangle$ acts on the initial state $\mid\psi_{in}\rangle$  as:

\begin{multline}
s^i\otimes s^j \otimes s^k\frac{1}{\sqrt{3}}\left(|000\rangle+|111\rangle+|222\rangle\right)= \\
= \frac{1}{\sqrt{3}}\left(|0+i\;0+j\;0+k\rangle+|1+i\;1+j\;1+k\rangle+|2+i\;2+j\;2+k\rangle\right).
\end{multline}
Note that the superscripts denotes powers of the generator and that the addition is modulo 3. In the case under study, where there is no preference profile over the different choices, any combination of the operators in $S=\{s^0,s^1,s^2\}$ leads to the same payoffs when applied to $\mid\psi_{in}\rangle$ as to $|000\rangle$.
We form a density matrix $\rho_{in}$  out of the initial state $\mid\psi_{in}\rangle$ and add noise that can be controlled by the parameter $f$ \cite{schmid}. We get:
\begin{equation}
\rho_{in}=f\mid\psi_{in}\rangle\langle\psi_{in}\mid+\frac{1-f}{27}I_{27},
\end{equation}
 where $I_{27}$ is the $27\times 27$ identity matrix. Alice, Bob and Charlie now applies a unitary operator $U$ that maximizes their chances of receiving a payoff $\$ =1$, and thereby the initial state $\rho_{in}$  is transformed into the final state $\rho_{fin}$.
\begin{equation}
\rho_{fin}=U\otimes U\otimes U \rho_{in} U^{\dagger}\otimes U^{\dagger}\otimes U^{\dagger}.
 \end{equation}
 We define for each player $i$  a payoff-operator $P_{i}$ , which contains the sum of orthogonal projectors associated with the states for which player $i$  receives a payoff $\$=1$ . For Alice this would correspond to
 \begin{multline}
 P_{A}=\left(\sum_{x_3,x_2,x_1=0}^{2}|x_{3}x_{2}x_{1}\rangle\langle x_{3}x_{2}x_{1}|,\, x_3\neq x_2, x_3\neq x_1, x_2\neq x_1\right)+\\
 +\left(\sum_{x_3,x_2,x_1=0}^{2}|x_{3}x_{2}x_{1}\rangle\langle x_{3}x_{2}x_{1}|,\, x_3=x_2\neq x_1\right).
 \end{multline}
 The expected payoff $E_{i}(\$)$  of player $i$  is as usual calculated by taking the trace of the product of the final state $\rho_{fin}$  and the payoff-operator $P_{i}$:
\begin{equation}
E(\$_{i})=\mathrm{Tr\left(\mathit{P}_{i}\rho_{fin}\right)}.
\end{equation}
It can be shown that if Alice, Bob and Charlie acts with a general SU(3), there exist a $U^{opt}(\phi,\theta^,\chi^,\alpha_1,\alpha_2,\alpha_3,\beta_1,\beta_2)\in \textrm{SU(3)}$, given in table 2, that outperforms classical randomization. 

\begin{table}\center
\label{Uopttable}
\begin{tabular}{|c|c|c|c|c|c|c|c|c|c|}
\hline
\hline
 Parameter  & $\phi$ & $\theta$  & $\chi$ & $\alpha_1$ & $\alpha_2$ & $\alpha_3$ & $\beta_1$ & $\beta_2$\tabularnewline

\hline
 Value  & $\frac{\pi }{4}$ & $\cos ^{-1}\left(\frac{1}{\sqrt{3}}\right)$ & $\frac{\pi }{4}$ & $\frac{5 \pi }{18}$ & $\frac{5 \pi }{18}$ & $\frac{5 \pi }{18}$ & $\frac{\pi }{3}$ & $\frac{11 \pi }{6}$ \tabularnewline
\hline
\end{tabular}
\caption{$U^{opt}$ in the given parametrization.}
\end{table}

The strategy profile $U^{opt}\otimes U^{opt}\otimes U^{opt}$ leads to a payoff of $E(\$)=\frac{6}{9}$, assuming ($f=1$), compared to the classical $E^c(\$)=\frac{4}{9}$. Letting the payoff function depend on the fidelity parameter $f$, we get a payoff function $E(\$(f))=\frac{2}{9}(f+2)$ where we can clearly see that the expected payoff reaches the classical value as $f \rightarrow 0$.

\section{Outlook}
As the field of quantum information theory matures and information processing moves into the quantum realm, will it be increasingly important to study the broad spectrum of effects of this transition. Game theory is the study of strategic decision making under limited information. How decision making should or will change as situations are played out in a world where this information is \emph{quantum} information, will be some of many conceptual challenges to address if classical communication and computing, is due to be replaced by systems governed by the peculiar and counter-intuitive laws of quantum mechanics.

\paragraph{\textbf{Acknowledgments:}}  The  work was supported  by the Swedish Research Council (VR).

\end{document}